\documentclass[aps,prb,preprint,showpacs,eqsecnum]{revtex4-1}
\usepackage{graphicx}
\usepackage{bm}
\usepackage{amsmath,upgreek}
\usepackage{txfonts}
\usepackage{hyperref}
\hypersetup{colorlinks=true,allcolors=blue}

\begin{document}
\title{Dissipation-induced rotation of suspended ferromagnetic nanoparticles}
\author{T.~V.~Lyutyy$^{1}$}
\email{lyutyy@oeph.sumdu.edu.ua}
\author{S.~I.~Denisov$^{1}$}
\email{denisov@sumdu.edu.ua}
\author{P.~H\"{a}nggi$^{2}$}
\email{hanggi@physik.uni-augsburg.de}
\affiliation{$^{1}$Sumy State University, 2 Rimsky-Korsakov Street, UA-40007 Sumy, Ukraine\\
$^{2}$Institut f\"{u}r Physik, Universit\"{a}t Augsburg, Universit\"{a}tsstra{\ss}e 1, D-86135 Augsburg, Germany}


\begin{abstract}
We report the precessional rotation of magnetically isotropic ferromagnetic nanoparticles in a viscous liquid that are subjected to a rotating magnetic field. In contrast to magnetically anisotropic nanoparticles, the rotation of which occurs due to coupling between the magnetic and lattice subsystems through magnetocrystalline anisotropy, the rotation of isotropic nanoparticles is induced only by magnetic dissipation processes. We propose a theory of this phenomenon based on a set of equations describing the deterministic magnetic and rotational dynamics of such particles. Neglecting inertial effects, we solve these equations analytically, find the magnetization and particle precessions in the steady state, determine the components of the particle angular velocity and analyze their dependence on the model parameters. The possibility of experimental observation of this phenomenon is also discussed.
\end{abstract}
\maketitle

\section{INTRODUCTION}
\label{Intr}

The ferromagnetic single-domain nano\-par\-ticles are objects of intense research mainly because of their high potential for various applications. Among them, the most promising are biomedical applications such as magnetic particle imaging \cite{Gleich2005, Pankhurst_2009}, drug delivery \cite{ARRUEBO200722, chemrev.5b00589}, magnetic fluid hyper\-thermia \cite{LAURENT20118, doi:10.1063/1.4935688} and cell separation \cite{Inglis2004, C4LC01422G, C8AN01061G}. The development of methods for the production of ferromagnetic nanoparticles with specified properties is one of the key factors for the realization of these and other applications. Up to date, a number of such methods have already been proposed and demonstrated (see, e.g., Refs.\ \cite{adma.201000260, YELENICH2014129, zhu7b00407} and references therein). Another key factor is the development of theoretical approaches aimed at a more complete description of the magnetic properties of ferromagnetic particles in viscous liquids subjected to external magnetic fields.

These systems are often studied in the framework of the rigid dipole model, when the particle magnetization is assumed to be directed along the particle easy axis. This approximation, which holds if the anisotropy magnetic field is large enough, was used to study, e.g., the effects of particle rotation, dipolar interaction and thermal fluctuations \cite{PhysRevE.79.021407, PhysRevE.83.021401, POLYAKOV20131483, PhysRevE.92.042312, Lyutyy2018a}. The same approximation was also used to describe in an analytical way the directed transport of suspended ferromagnetic nanoparticles induced by the Magnus force \cite{DenPed2017, DENISOV201789, PhysRevE.97.032608}.

However, if the anisotropy magnetic field does not strongly exceed the external field, then the model of suspended particles with ``frozen" magnetization (i.e., the rigid dipole model) fails. In this case, it is necessary to consider a coupled rotation of the particle magnetization and particle body. One of the most simple and effective methods to determine properties of such systems is based on the concept of relaxation times (see, e.g., Ref.\ \cite{ROSENSWEIG2002370}). At the same time, the dynamical approach based on both deterministic and stochastic equations of motion for the particle and its magnetization provides a much more complete description of the system's properties. Although these equations were derived many years ago \cite{Cebers1975}, only recently they have been rederived and applied for studying the coupling between the magnetic and rotational dynamics of suspended particles and effects in magnetic fluid hyperthermia \cite{USOV2015339, Usadel2015, PhysRevB.95.104430, PhysRevB.95.134447, LYUTYY201887, LYUTYY2019198}.

In this paper, we use the dynamical approach to obtain two main results for magnetically isotropic (i.e.,  without magnetocrystalline anisotropy) ferromagnetic nanoparticles in a viscous liquid. First, there exists the dynamical coupling between the magnetic and lattice subsystems in such particles arising from magnetic dissipation. Second, due to this coupling, a rotating magnetic field induces the precessional rotation of these particles. We hope that these rather surprising results will stimulate experimental studies in this area.

The paper is structured as follows. In Sec.\ \ref{Basic}, we introduce the basic equations describing the coupled magnetic and rotational dynamics of ferromagnetic nanoparticles in a viscous liquid. The case of magnetically isotropic nanoparticles subjected to a rotating magnetic field is considered in Sec.\ \ref{Rot}. Here, assuming that inertial effects are negligible, we solve these equations in the steady state, analyze stability of obtained solutions and investigate, both analytically and numerically, the dependence of the magnetization precession and precessional rotation of nanoparticles on the model parameters. Our main conclusions are summarized in Sec.\ \ref{Concl}.

\section{Basic equations for coupled magneto-mechanical dynamics}
\label{Basic}

We consider single-domain particles of sphe\-rical form suspended in a viscous liquid and characterized by the total momentum $\mathbf{J} = \mathbf{J}(t)$ defined as a sum of the mechanical angular momentum, $I \boldsymbol {\omega}$, and the spin momentum in the quasi-classical approximation, $- (V/\gamma) \mathbf{M}$:
\begin{equation}
    \mathbf{J} = I \boldsymbol {\omega}
    - \frac{V}{\gamma}\mathbf{M}.
    \label{Def_J}
\end{equation}
Here, $I= \rho_{m} V d^{2}/10$ is the particle's moment of inertia, $\rho_{m}$, $V$ and $d$ are the particle density, volume and diameter, respectively, $\boldsymbol {\omega} = \boldsymbol {\omega}(t)$ is the particle angular velocity, $\mathbf{M} = \mathbf{M}(t)$ ($|\mathbf{M}| = M = \mathrm{const}$) is the particle magnetization, and $\gamma(>0)$ is the gyromagnetic ratio. In the absence of dissipation we have \cite{USOV2015339, Usadel2015} $d\mathbf{J} /dt = V\mathbf{M} \times \mathbf{H}$, where $\mathbf{H} = \mathbf{H}(t)$ is an external magnetic field and the sign $\times$ denotes the vector product. Therefore, by differentiating the particle angular momentum $I \boldsymbol {\omega}$ with respect to time and introducing the frictional torque $-6\eta V \boldsymbol {\omega}$ ($\eta$ is the dynamic viscosity of the liquid) acting on the particle, one obtains the equation
\begin{equation}
    I\frac{d}{dt} \boldsymbol {\omega} =
    \frac{V}{\gamma} \frac{d}{dt} \mathbf{M}
    + V \mathbf{M} \times \mathbf{H} -
    6\eta V \boldsymbol {\omega}
    \label{dyn_omega}
\end{equation}
describing the rotation of ferromagnetic particles in a viscous liquid. Note that for vacuum (when $\eta = 0$) this equation was introduced in Ref.\ \cite{USOV2015339}, and for a viscous liquid it was introduced in Ref.\ \cite{Usadel2015}.

Because the magnetization value $M$ is assumed to be time-independent, the dynamics of the magnetization vector $\mathbf{M}$ can be described, e.g., by the Landau-Lifshitz (LL) equation \cite{Bertotti2009}. An important feature of this equation describing the magnetization dynamics in rotating nanoparticles is that its dissipation term should be properly modified \cite{USOV2015339}:
\begin{equation}
    \frac{d}{dt} \mathbf{M} = -\gamma
    \mathbf{M} \times \mathbf{H}_{\textrm
    {eff}} - \frac{\gamma \alpha}{M}
    \mathbf{M} \times \left(\mathbf{M}\times
    \left(\mathbf{H}_{\textrm{eff}} - \frac{ \boldsymbol{\omega}}{\gamma}\right)
    \right),
    \label{LLeq}
\end{equation}
where $\alpha (>0)$ is the LL damping parameter, $-\boldsymbol{ \omega} /\gamma$ is the so-called Barnett field originating from the particle rotation (see, e.g., Ref.\ \cite{PhysRevB.95.134447}), and $\mathbf{H}_{ \textrm {eff}}$ is the effective magnetic field acting on $\mathbf{M}$. In particular, in the case of uniaxial particles the effective magnetic field is given by
\begin{equation}
    \mathbf{H}_{\textrm {eff}} = \frac
    {H_{a}}{M}(\mathbf{M}\cdot\mathbf{n})
    \mathbf{n} + \mathbf{H}.
    \label{Heff}
\end{equation}
Here, $H_{a}$ is the uniaxial magnetic anisotropy field, the dot denotes the scalar product, and the unit vector $\mathbf{n}$ is directed along the easy axis of magnetization and satisfies the following equation of motion (kinematic relation):
\begin{equation}
    \frac{d}{dt} \mathbf{n} =
    \boldsymbol {\omega}\times \mathbf{n}.
    \label{dyn_n}
\end{equation}

Equations \textcolor{blue}{(\ref{dyn_omega})}, \textcolor{blue}{ (\ref{LLeq})} and \textcolor{blue}{ (\ref{dyn_n})} supplemented by the effective magnetic field \textcolor{blue}{ (\ref{Heff})} completely describe the coupled dynamics of magnetization and rotational dynamics of uniaxial nanoparticles in a viscous liquid. Introducing the dimensionless variables and parameters
\begin{align}
    &\mathbf{m} = \frac{\mathbf{M}}{M},\;
    \boldsymbol {\nu} = \frac{\boldsymbol
    {\omega}}{\gamma M},\; \mathbf{h}_{
    \textrm{eff}}= \frac{\mathbf{H}_{
    \textrm{eff}}}{M},\; \mathbf{h} =
    \frac{\mathbf{H}}{M}, \nonumber
    \\[2pt]
    &\tau = \gamma Mt,\; \kappa = \frac{I
    \gamma^2}{V},\; \beta = \frac{6\gamma
    \eta}{M},\; h_{a} = \frac{H_{a}}{M},
\label{DimLess}
\end{align}
these equations can be rewritten in the dimensionless form
\begin{subequations}\label{Dyn:LL}
\begin{gather}
    \kappa\dot{\boldsymbol {\nu}} = \dot{
    \mathbf{m}} + \mathbf{m} \times \mathbf{h}
    - \beta \boldsymbol {\nu},
    \label{Dyn:LL1}\\[2pt]
    \qquad \dot{\mathbf{m}} = - \mathbf{m}
    \times \mathbf{h}_{\textrm{eff}} -\alpha
    \mathbf{m} \times \left(\mathbf{m}\times
    (\mathbf{h}_{\textrm{eff}} - \boldsymbol
    {\nu}) \right),
    \label{Dyn:LL2}\\[2pt]
    \dot{\mathbf{n}} = \boldsymbol {\nu}
    \times \mathbf{n},
    \label{Dyn:LL3}
\end{gather}
\end{subequations}
where the overdot denotes the derivative with respect to the dimensionless time $\tau$ and, according to \textcolor{blue}{ (\ref{Heff})},
\begin{equation}
    \mathbf{h}_{\textrm{eff}} = h_{a}(\mathbf{m}
    \cdot \mathbf{n})\mathbf{n} + \mathbf{h}.
    \label{h_eff}
\end{equation}

If the magnetization dynamics is assumed to be governed by the Landau-Lifshitz-Gilbert (LLG) equation \cite{Gilbert2004}, then, to take into account  the influence of particle rotation, this equation should be modified as follows \cite{Usadel2015}:
\begin{equation}
    \frac{d}{dt} \mathbf{M} = -\gamma
    \mathbf{M} \times \mathbf{H}_{\textrm
    {eff}} + \frac{\alpha'}{M} \mathbf{M}
    \times \left(\frac{d}{dt} \mathbf{M} -
    \boldsymbol{\omega} \times \mathbf{M}
    \right)
    \label{LLGeq}
\end{equation}
($\alpha'$ is the LLG damping parameter). Using the relation
\begin{equation*}
    \mathbf{M} \times \frac{d}{dt}\mathbf{M}
    = -\gamma \mathbf{M} \times (\mathbf{M}
    \times \mathbf{H}_{\textrm{eff}}) -
    \alpha'M \frac{d}{dt}\mathbf{M} - \alpha'M
    (\mathbf{M} \times \boldsymbol{\omega})
    \label{rel1}
\end{equation*}
that follows directly from \textcolor{blue}{(\ref{LLGeq})} and notations \textcolor{blue}{(\ref{DimLess})}, Eq.\ \textcolor{blue}{ (\ref{LLGeq})} can be reduced to the dimensionless LLG equation
\begin{equation}
    (1+\alpha'^{2}) \dot{\mathbf{m}} = -
    \mathbf{m} \times (\mathbf{h}_{\textrm{eff}}
    + \alpha'^{2} \boldsymbol{\nu}) - \alpha'
    \mathbf{m} \times \left(\mathbf{m}\times
    (\mathbf{h}_{\textrm{eff}} - \boldsymbol
    {\nu}) \right),
    \label{Dyn:LLG}
\end{equation}
which in this case should be used instead of Eq.\ \textcolor{blue}{ (\ref{Dyn:LL2})}.

By comparing Eqs.\ \textcolor{blue}{ (\ref{Dyn:LL2})} and \textcolor{blue}{ (\ref{Dyn:LLG})}, we can make sure that at $\boldsymbol {\nu} = \mathbf{0}$ (when nanoparticles do not move) these equations are, in fact, equivalent \cite{Bertotti2009}. Strictly speaking, at $\boldsymbol {\nu} \neq \mathbf{0}$ these equations are different. However, in the most common case, when $\alpha' \ll 1$, this difference can be neglected. Therefore, in further analysis we will use the set of Eqs.\ \textcolor{blue}{ (\ref{Dyn:LL})}.

It is important to emphasize that Eqs.\ \textcolor{blue}{ (\ref{Dyn:LL})} are written in the deterministic approximation. In principle, thermal fluctuations can also be accounted for by introducing in these equations the Gaussian white noises (see, e.g., Refs.\ \cite{Cebers1975, Usadel2015, PhysRevB.95.104430}). However, if these noises are not too strong, they do not destroy the deterministic effects. This means that Eqs.\ \textcolor{blue}{ (\ref{Dyn:LL})} can be used as a starting point for studying the coupled magnetic and rotational dynamics of suspended ferromagnetic particles. It is also possible to formulate conditions under which thermal fluctuations can safely be neglected. In particular, the magnetization fluctuations in magnetically isotropic nanoparticles (when $h_{a} = 0$) may be considered as small if the magnetic energy $MHV$ exceeds the thermal energy $k_{B}T$ ($k_{B}$ is the Boltzmann constant and $T$ is the absolute temperature). This situation occurs when the particle diameter satisfies the condition $d > d_{1}$, where $d_{1} = (6k_{B}T /\pi MH)^{1/3}$ (see also Sec.\ \ref{NumRes}).

\section{Precessional rotation of magnetically isotropic nanoparticles}
\label{Rot}

Our next aim is to study the coupled magnetic and rotational dynamics of isotropic ferromagnetic nanoparticles. Since in this case $h_{a} = 0$ and, according to \textcolor{blue}{ (\ref{h_eff})}, $\mathbf{h}_{\textrm{eff}} = \mathbf{h}$, this dynamics is described by a set of only two equations, \textcolor{blue}{ (\ref{Dyn:LL1})} and \textcolor{blue}{ (\ref{Dyn:LL2})}. As to Eq.\ \textcolor{blue}{ (\ref{Dyn:LL3})}, for these nanoparticles it can be excluded from further consideration.

Assuming that the left-hand side of Eq.\ \textcolor{blue}{ (\ref{Dyn:LL1})} is negligibly small (for more detail, see Sec.\ \ref{NumRes}), we can rewrite the set of Eqs.\ \textcolor{blue}{ (\ref{Dyn:LL1})} and \textcolor{blue}{ (\ref{Dyn:LL2})} in the form
\begin{subequations}\label{Dyn:LL_I}
\begin{gather}
    \dot{\mathbf{m}} = -\mathbf{m} \times
    \mathbf{h} + \beta \boldsymbol {\nu},
    \label{Dyn:LL1_I}\\[2pt]
    \qquad \dot{\mathbf{m}} = - \mathbf{m}
    \times \mathbf{h} -\alpha \mathbf{m}
    \times \left(\mathbf{m}\times
    (\mathbf{h} - \boldsymbol{\nu}) \right).
    \label{Dyn:LL2_I}
\end{gather}
\end{subequations}
Substituting $\dot{\mathbf{m}}$ from Eq.\ \textcolor{blue}{ (\ref{Dyn:LL2_I})} into Eq.\ \textcolor{blue}{ (\ref{Dyn:LL1_I})} and taking into account that, according to \textcolor{blue}{ (\ref{Dyn:LL1_I})}, $\boldsymbol{\nu} \cdot \mathbf{m} = 0$, one obtains
\begin{equation}
    \boldsymbol{\nu} = - \frac{\alpha}
    {\alpha + \beta} \mathbf{m} \times
    (\mathbf{m}\times \mathbf{h}).
    \label{nu1}
\end{equation}
Then, substituting this expression for the dimensionless angular velocity into Eq.\ \textcolor{blue}{ (\ref{Dyn:LL2_I})}, it is not difficult to derive a closed LL equation for the unit magnetization vector of a magnetically isotropic particle in a viscous liquid
\begin{equation}
    \dot{\mathbf{m}} = - \mathbf{m}
    \times \mathbf{h} - q \mathbf{m}
    \times (\mathbf{m}\times \mathbf{h}),
    \label{LL1}
\end{equation}
where $q = \alpha \beta/(\alpha + \beta)$. Note that the limit $\beta \to \infty$ corresponds to immobile particles. In this limit, $\boldsymbol {\nu} \to \mathbf{0}$, $q \to \alpha$, and Eq.\ \textcolor{blue}{ (\ref{LL1})} reduces to the standard LL equation.

Next, we use Eq.\ \textcolor{blue}{ (\ref{LL1})} and expression \textcolor{blue}{ (\ref{nu1})} to study the magnetic and rotational steady-state dynamics of isotropic nanoparticles subjected to the rotating magnetic field
\begin{equation}
    \mathbf{h} = h(\cos{\upsilon \tau}\,
    \mathbf{e}_{x} + \rho\sin{\upsilon
    \tau}\, \mathbf{e}_{y}).
    \label{h}
\end{equation}
Here, $h = |\mathbf{h}| = \mathrm{const}$ is the dimensionless amplitude of the rotating magnetic field, $\upsilon = \Omega/\gamma M$,  $\Omega$ is the rotating field frequency, $\rho = \pm 1$ is the parameter that determines the direction of the magnetic field rotation, and $\mathbf{e}_{x}$, $\mathbf{e}_{y}$ and $\mathbf{e}_{z}$ are the unit vectors of the Cartesian coordinate system.

\subsection{Magnetization precession}
\label{MagnPr}

Let us represent the unit magnetization vector $\mathbf{m}$ in the form
\begin{equation}
    \mathbf{m} = \sin{\theta} \cos{\varphi}
    \, \mathbf{e}_{x} + \sin{\theta}
    \sin{\varphi}\,\mathbf{e}_{y} + \cos{\theta}
    \,\mathbf{e}_{z},
    \label{def_m}
\end{equation}
where $\theta = \theta(\tau)$ and $\varphi = \varphi(\tau)$ are the polar and azimuthal angles of $\mathbf{m}$, respectively. Then, introducing the lag angle
\begin{equation}
    \psi = \rho\upsilon \tau - \varphi,
    \label{def_psi}
\end{equation}
we can reduce the vector LL equation \textcolor{blue}{ (\ref{LL1})} to a set of differential equations for $\theta$ and $\psi$
\begin{align}
    &\dot{\theta} = h\sin{\psi} + qh
    \cos{\theta} \cos{\psi},\nonumber
    \\[2pt]
    &(\dot{\psi} - \rho \upsilon)
    \sin{\theta} = h\cos{\theta}
    \cos{\psi} -qh\sin{\psi}.
\label{DiffEqs}
\end{align}

Assuming that in the steady state (when $\tau \to \infty$) the angles $\theta$ and $\psi$ do not depend on time,
\begin{equation}
    \theta = \theta_{\rho} = \mathrm{const},
    \quad \psi = \psi_{\rho} = \mathrm{const}
    \label{theta,psi}
\end{equation}
($0 \leq \theta_\rho \leq \pi$, $-\pi < \psi_\rho \leq \pi$), from Eqs.\ \textcolor{blue}{(\ref{DiffEqs})} one gets a set of equations for  $\theta_\rho$ and $\psi_\rho$
\begin{align}
    &\sin{\psi_\rho} + q\cos{\theta_\rho}
    \cos{\psi_\rho} = 0, \nonumber
    \\[2pt]
    &\rho\chi \sin{\theta_\rho} + \cos{
    \theta_\rho} \cos{\psi_\rho} - q\sin{
    \psi_\rho} = 0,
\label{Eqs_theta,psi_rho}
\end{align}
where $\chi = \upsilon/h$. If these angles are represented in the form
\begin{equation}
    \theta_\rho = \frac{\pi}{2}(1+\rho)
    - \rho \theta_0, \quad
    \psi_\rho = \rho \psi_0,
    \label{theta,psi_rho}
\end{equation}
then new variables $\theta_0$ ($0 \leq \theta_0 \leq \pi$) and $\psi_0$ ($-\pi < \psi_0 \leq \pi$) do not depend on the parameter $\rho$. Indeed, taking into account that $\sin{\psi_\rho} = \rho \sin{\psi_0}$, $\cos{\psi_\rho} = \cos{\psi_0}$, $\sin{\theta_\rho} = \sin{\theta_0}$ and $\cos{\theta_\rho} = -\rho \cos{\theta_0}$, the set of Eqs.\ \textcolor{blue} {(\ref{Eqs_theta,psi_rho})} readily yields
\begin{align}
    &\sin{\psi_0} - q\cos{\theta_0}
    \cos{\psi_0} = 0, \nonumber
    \\[2pt]
    &\chi q\sin{\theta_0} - (1+q^2)
    \sin{\psi_0} = 0.
\label{Eqs_theta,psi_0}
\end{align}

According to the last equation in \textcolor{blue} {(\ref{Eqs_theta,psi_0})}, the angle $\psi_0$ (like $\theta_0$) must belong to the interval $[0,\pi]$, i.e., only non-negative values of $\sin{\psi_0}$ are permitted. Introducing parameters $c = q/\!\sqrt{1+q^2}$ and $k = \chi/\! \sqrt{1+q^2}$, it can be easily shown from Eqs.\ \textcolor{blue} {(\ref{Eqs_theta,psi_0})} that $\sin{\psi_0}$ satisfies the biquadratic equation
\begin{equation}
    \sin^{4}{\psi_0} - (1 + k^2)\sin^{2}
    {\psi_0} + c^2 k^2 = 0.
    \label{Eq_psi0}
\end{equation}
Since $\sin{\psi_0} \in [0,1]$, its unique solution is given by $\sin{\psi_0} = R$, where
\begin{equation}
    R = \frac{1}{\sqrt{2}}\! \sqrt{
    1 + k^2 - \sqrt{(1 + k^{2})^2 -
    4c^2 k^2 } }.
    \label{defR}
\end{equation}
From this, using the second equation in \textcolor{blue} {(\ref{Eqs_theta,psi_0})}, one obtains $\sin{\theta_0} = R/ck$ [note that, according to \textcolor{blue} {(\ref{defR})}, the conditions $R \leq 1$ and $R/ck \leq 1$ always hold]. In addition, the first equation in \textcolor{blue} {(\ref{Eqs_theta,psi_0})} shows that both angles $\psi_0$ and $\theta_0$ must lie either in the interval $[0, \pi/2)$ or in the interval $(\pi/2, \pi]$. In the former case, the solution of Eqs.\ \textcolor{blue} {(\ref{Eqs_theta,psi_0})} is written as
\begin{equation}
    \theta^{(1)}_0 = \arcsin{\frac{R}{ck}},
    \quad
    \psi^{(1)}_0 = \arcsin{R},
    \label{Sol0_1}
\end{equation}
while in the latter case it is written as $\theta^{(2)}_0 = \pi - \theta^{(1)}_0$, $\psi^{(2)}_0 = \pi - \psi^{(1)}_0$.

Thus, the rotating magnetic field \textcolor{blue} {(\ref{h})} could, in principle, induce in magnetically isotropic nanoparticles two steady-state precessional states of the magnetization, $\mathbf{m}^{(1)}$ and $\mathbf{m}^{(2)}$. Using \textcolor{blue} {(\ref{theta,psi_rho})} and \textcolor{blue} {(\ref{Sol0_1})}, we find the angles
\begin{equation}
    \theta^{(1)}_{\rho} = \frac{\pi}{2}
    (1+\rho) - \rho\arcsin{\frac{R}{ck}},
    \quad
    \psi^{(1)}_{\rho} = \rho \arcsin{R}
    \label{Sol_rho_1}
\end{equation}
for $\mathbf{m}^{(1)}$, and $\theta^{(2)}_{\rho} = \theta^{(1)}_{-\rho}$ and $\psi^{(2)}_\rho = \rho \pi + \psi^{(1)}_{-\rho}$ for $\mathbf{m}^{(2)}$. These expressions, together with the definition \textcolor{blue} {(\ref{def_psi})} of the lag angle, allow us to determine the components of the vector $\mathbf{m}^{(l)}$ ($l = 1, 2$) as follows:
\begin{equation}
    \begin{pmatrix}
    m_{x}^{(l)} \\[4pt] m_{y}^{(l)}
    \\[4pt] m_{z}^{(l)}
    \end{pmatrix} = (-1)^{1+l} \frac{R}{ck}
    \times \left\{\!\!
    \begin{array}{ll}
    \vphantom{m_{x}^{(l)}}
    \cos{\left(\upsilon \tau + (-1)^l
    \arcsin{R}\right)},
    \\ [4pt]
    \vphantom{m_{y}^{(l)}}
    \rho \sin{\left(\upsilon\tau + (-1)^l
    \arcsin{R}\right)},
    \\[4pt]
    \vphantom{m_{z}^{(l)}}
    \! -\rho \sqrt{c^2k^2/R^2 - 1}.
    \end{array}
    \right.
    \label{m_(l)}
\end{equation}
According to them, the steady-state magnetization precessions, if they are stable, occur about the $z$ axis with the magnetic field frequency and their direction coincides with the direction of the magnetic field rotation. The time-averaged magnetization in these precessional states, defined as $\langle \mathbf{m}^{(l)} \rangle = (\upsilon /2\pi) \int_{0}^{2\pi/\upsilon}\! \mathbf{m}^{(l)}d\tau$, is given by $\langle \mathbf{m}^{(l)} \rangle = (-1)^l \rho \sqrt{1- R^2/c^2 k^2}\, \mathbf{e}_z$, i.e., the magnetic field rotating in the $xy$ plane magnetizes the isotropic nanoparticles in the $z$ direction. We note in this context that a similar effect, the magnetization of nanoparticle systems by a rotating magnetic field, was earlier predicted and analyzed for anisotropic (uniaxial) and immobile nanoparticles \cite{Denisov2006, Denisov_2006, PhysRevB.75.184432}. But its nature is quite different: the magnetization of those systems occurs due to the presence of the anisotropy magnetic field, which in our case is absent.

\subsection{Stability analysis of the magnetization precession}
\label{Stab}

Now we analyze the linear stability of the precessional states $\mathbf{m}^{(l)}$. Substituting $\theta = \theta_{\rho}^{(l)} + \theta_1$ ($\theta_1 = \theta_1(\tau)$, $|\theta_1| \ll 1$) and $\psi = \psi_{\rho}^ {(l)} + \psi_1$ ($\psi_1 = \psi_1(\tau)$, $|\psi_1| \ll 1$) into Eqs.\ \textcolor{blue} {(\ref{DiffEqs})}, we obtain a set of ordinary differential equations for $\theta_1$ and $\psi_1$
\begin{align}
    &\dot{\theta}_1 = h\big(\!\cos{
    \psi_\rho^{(l)}} - q\cos{\theta_
    \rho^{(l)}} \sin{\psi_{\rho}^{(l)}}
    \big)\,\psi_1 -qh
    \sin{\theta_\rho^{(l)}}\cos{
    \psi_\rho^{(l)}}\,\theta_1,
    \nonumber \\[2pt]
    &\dot{\psi}_1\sin{\theta_\rho^{(l)}} =
    - h\big(\!\cos{\theta_\rho^{(l)}} \sin{
    \psi_\rho^{(l)}} + q\cos{\psi_\rho^{(l)}}
    \big)\,\psi_1 + h\big(\rho\chi\cos{\theta_
    \rho^{(l)}}
    \nonumber \\[2pt]
    & \phantom{\dot{\psi}_1\sin{\theta_
    \rho^{(l)}} =} - \sin{\theta_\rho^{(l)}}
    \cos{\psi_\rho^{(l)}}\big)\,\theta_1.
    \label{Eqs_theta1,psi1}
\end{align}
Assuming that
\begin{equation}
    \theta_1 = \tilde{\theta}_1 e^{\lambda_l
    h\tau}, \quad
    \psi_1 = \tilde{\psi}_1 e^{\lambda_l
    h\tau},
    \label{def_theta1,psi1}
\end{equation}
where the parameters $\tilde{\theta}_1$, $\tilde{\psi}_1$ and $\lambda_l$ do not depend on $\tau$, Eqs.\ \textcolor{blue} {(\ref{Eqs_theta1,psi1})} are reduced to a homogeneous system of linear equations with respect to $\tilde{\theta}_1$ and $\tilde{\psi}_1$, which can be written in the matrix form as
\begin{equation}
    \begin{pmatrix}
    \lambda_l + (-1)^{1+l}a_{11} &
    (-1)^{l}a_{12} \\[4pt]
    (-1)^{1+l}a_{21} & \lambda_l +
    (-1)^{l}a_{22}
    \end{pmatrix}
    \begin{pmatrix}
    \tilde{\theta}_1 \\[4pt]
    \tilde{\psi}_1
    \end{pmatrix} =
    \begin{pmatrix}
    0 \\[4pt] 0
    \end{pmatrix}.
    \label{theta1,psi1}
\end{equation}
Taking into account that, according to \textcolor{blue} {(\ref{theta,psi_rho})} and \textcolor{blue} {(\ref{Sol_rho_1})},
\begin{align}
    &\sin{\theta_{\rho}^{(l)}} = \frac{R}
    {ck}, \quad \cos{\theta_{\rho}^{(l)}} =
    (-1)^{l}\rho \frac{1}{ck} \sqrt{c^2k^2 -
    R^2},
    \nonumber \\[2pt]
    &\sin{\psi_{\rho}^{(l)}} = \rho R,
    \quad \cos{\psi_{\rho}^{(l)}} =
    (-1)^{1+l}\sqrt{1-R^2},
\label{theta,psi_rho2}
\end{align}
the coefficients $a_{nm}$ in Eqs.\ \textcolor{blue} {(\ref{theta1,psi1})} are expressed through the parameters $c$ and $k$ (recall that $c<1$ and $k<\infty$) as follows:
\begin{align}
    &a_{11} = \frac{R}{k} \sqrt{\frac{1 -
    R^2}{1-c^2}},
    \nonumber \\
    &a_{12} = \sqrt{1-R^2} + \frac{R}{k}
    \sqrt{\frac{c^2k^2 - R^2}{1-c^2}},
    \nonumber \\
    &a_{21} = \sqrt{1-R^2} + \frac{k}{R}
    \sqrt{\frac{c^2k^2 - R^2}{1-c^2}},
    \nonumber \\
    &a_{22} = \sqrt{c^2k^2-R^2} - \frac{c^2k}
    {R} \sqrt{\frac{1 - R^2}{1-c^2}}.
\label{a_nm}
\end{align}

It is well known that non-zero solutions of the system of Eqs.\ \textcolor{blue} {(\ref{theta1,psi1})} exist only if the determinant of the coefficient matrix vanishes, i.e., if
\begin{equation}
    \lambda_{l}^2 + (-1)^{1+l}(a_{11} - a_{22})
    \lambda_l +a_{12}a_{21} - a_{11}a_{22} = 0.
    \label{Eq_lambda}
\end{equation}
This occurs at $\lambda_l = \lambda_{l}^{+}$ and $\lambda_l = \lambda_{l}^{-}$, where
\begin{equation}
    \lambda_{l}^{\pm} = (-1)^{l} \frac{1}{2}
    (a_{11} - a_{22}) \pm i \frac{1}{2}\!
    \sqrt{4a_{12}a_{21} - (a_{11} + a_{22})^2}
    \label{lambda_pm}
\end{equation}
($i$ is the imaginary unit) are solutions of Eq.\ \textcolor{blue} {(\ref{Eq_lambda})}. It can be verified (analytically or numerically) that $a_{11} - a_{22} > 0$ and $4a_{12}a_{21} - (a_{11} + a_{22})^2 > 0$ for all values of $c$ and $k$. Therefore, using \textcolor{blue} {(\ref{lambda_pm})}, one can conclude that the steady-state precessional state of the magnetization with $l=1$ (i.e., $\mathbf{m}^{(1)}$) is stable (because $\mathrm{Re}\, \lambda_{1}^{\pm} <0$), while the precessional state with $l=2$ (i.e., $\mathbf{m}^{(2)}$) is unstable (because $\mathrm{Re}\, \lambda_{2}^{\pm} >0$). Note also that, according to \textcolor{blue} {(\ref{lambda_pm})}, the magnetization approaches the stable steady state $\mathbf{m}^{(1)}$ in an oscillatory manner.

\subsection{Particle precession}
\label{PartPr}

Since the steady-state magnetization $\mathbf{m}^{(2)}$ is unstable, we only determine the components of the (dimensionless) particle angular velocity $\boldsymbol{\nu}$ that correspond to the stable steady-state magnetization $\mathbf{m}^{(1)}$. To this end, using \textcolor{blue} {(\ref{h})} and \textcolor{blue} {(\ref{nu1})}, we first represent the Cartesian components of $\boldsymbol{\nu}$ in the form
\begin{equation}
    \begin{pmatrix}
    \nu_{x} \\[5pt] \nu_{y}
    \\[5pt] \nu_{z}
    \end{pmatrix} = \frac{\alpha h}
    {\alpha+\beta}  \times \left\{\!\!
    \begin{array}{ll}
    \vphantom{\nu_{x}^{(l)}}
    \cos{\upsilon \tau} - m_{x}^{(1)}
    \left(m_{x}^{(1)} \cos{\upsilon
    \tau} + \rho m_{y}^{(1)} \sin{
    \upsilon \tau}\right),
    \\ [4pt]
    \vphantom{\nu_{x}^{(1)}}
    \rho \sin{\upsilon \tau} - m_{y}^{(1)}
    \left(m_{x}^{(1)} \cos{\upsilon\tau} +
    \rho m_{y}^{(1)} \sin{\upsilon\tau}
    \right),
    \\[4pt]
    \vphantom{\nu_{x}^{(1)}}
    \!- m_{z}^{(1)}\left(m_{x}^{(1)} \cos{
    \upsilon\tau} + \rho m_{y}^{(1)}
    \sin{\upsilon\tau}\right).
    \end{array}
    \right.
    \label{nu_2}
\end{equation}
Then, substituting the magnetization components $m_{x}^{(1)}$, $m_{y}^{(1)}$ and $m_{z}^{(1)}$ from \textcolor{blue} {(\ref{m_(l)})} into \textcolor{blue} {(\ref{nu_2})}, one straightforwardly obtains
\begin{align}
    &\nu_{x} = \frac{\alpha h}
    {\alpha+\beta} \Big[ \cos{\upsilon
    \tau} - \frac{R^2}{c^2k^2} \sqrt{
    1-R^2}
    \nonumber \\
    &\phantom{\nu_{x} =}
    \times \cos{\left(\upsilon \tau
    - \arcsin{R} \right)}\Big],
    \nonumber \\[2pt]
    &\nu_{y} = \rho \frac{
    \alpha h} {\alpha+\beta} \Big[
    \sin{\upsilon \tau} - \frac{R^2}
    {c^2k^2} \sqrt{1-R^2}
    \nonumber \\
    &\phantom{\nu_{y} =}
    \times \sin{\left(\upsilon \tau
    - \arcsin{R} \right)}\Big],
    \nonumber \\[2pt]
    &\nu_{z} = \rho \frac{
    \alpha h} {\alpha+\beta} \frac{
    R}{c^2k^2} \sqrt{(1-R^2)(c^2k^2-R^2)}.
\label{nu_3}
\end{align}
These expressions show that a rotating magnetic field causes a dissipation-induced precessional motion of magnetically isotropic nanoparticles. The precession occurs with the magnetic field frequency, the particle and magnetic field are rotated in the same direction, and the (dimensionless) magnitude of the particle angular velocity can be cast as
\begin{equation}
    |\boldsymbol{\nu}| = \frac{\alpha h}
    {\alpha + \beta} \frac{1}{ck} \sqrt{
    c^2k^2 - R^2 + R^4}.
    \label{|nu|}
\end{equation}
It should be also pointed out that, according to \textcolor{blue} {(\ref{m_(l)})} and \textcolor{blue} {(\ref{nu_3})}, $\boldsymbol{\nu} \cdot \mathbf{m}^{(1)} = 0$, i.e., the magnetization precession is synchronized with the particle rotation.

In order to get more insight into the dissipation-induced mechanism of nanoparticle rotation, we first determine the particle angular velocity in the cases of small fluid dynamic viscosity ($\beta \to 0$) and magnetic damping parameter ($\alpha \to 0$). Using \textcolor{blue} {(\ref{nu_3})} and \textcolor{blue} {(\ref{|nu|})}, one finds
\begin{equation}
    \nu_{z} = \rho \frac{h\chi}
    {1+\chi^2},
    \quad
    |\boldsymbol{\nu}| = \frac{h\chi}
    {\sqrt{1+\chi^2}}
    \label{beta_to_0}
\end{equation}
and
\begin{equation}
    \nu_{z} = \rho \frac{h\chi \alpha}
    {\beta(1+\chi^2)},
    \quad
    |\boldsymbol{\nu}| = \frac{h\chi \alpha}
    {\beta\sqrt{1+\chi^2}}
    \label{alpha_to_0}
\end{equation}
for the first and second cases, respectively. According to these results, the particle rotation exists at $\beta \to 0$, while it vanishes at $\alpha \to 0$. We therefore conclude that it is the magnetic dissipation that is responsible for the nanoparticle rotation.

Because the viscosity parameter $\beta$ is usually large (see below), it is also reasonable to consider the limiting case $\beta \to \infty$. In the main approximation in $1/\beta$, expressions \textcolor{blue} {(\ref{nu_3})} and \textcolor{blue} {(\ref{|nu|})} yield
\begin{equation}
    \nu_{z} = \rho\frac{h}{\sqrt{2}
    \beta b} \sqrt{a^2 -
    a\sqrt{a^2 - 4b^2}-2b^2}
    \label{nu_z2}
\end{equation}
and
\begin{equation}
    |\boldsymbol{\nu}| = \frac{h}
    {\sqrt{2}\beta} \sqrt{a -
    \sqrt{a^2 - 4b^2}},
    \label{|nu|2}
\end{equation}
where $a = 1 + \alpha^2 + \chi^2$ and $b = \alpha \chi$. They show that in this limit the nanoparticle angular velocity decreases to zero inversely proportional to the viscosity parameter. Note also that
\begin{equation}
    \nu_{z} = \rho \frac{qh\chi}
    {\beta(1+q^2)},
    \quad
    |\boldsymbol{\nu}| = \frac{qh\chi}
    {\beta\sqrt{1+q^2}}
    \label{chi_to_0}
\end{equation}
as $\chi \to 0$ and
\begin{equation}
    \nu_{z} = \rho \frac{qh}
    {\beta\chi},
    \quad
    |\boldsymbol{\nu}| = \frac{qh}
    {\beta}
    \label{chi_to_inf}
\end{equation}
as $\chi \to \infty$.

As it was mentioned in Sec.\ \ref{Basic}, a correct description of the magnetization dynamics in rotating nanoparticles is achieved by introducing in Eq.\ \textcolor{blue} {(\ref{LLeq})} the Barnett field. This emergent magnetic field is responsible for the Barnett effect (magnetization by rotation) \cite{PhysRev.6.239} and its existence has been recently confirmed experimentally for different spin systems \cite{Chudo_2014, wood_2017, PhysRevLett.122.177202}. In this context, it is of interest to analyze the role of the Barnett field in the dissipation-induced rotation of suspended ferromagnetic nanoparticles. In its absence, when the term $-\boldsymbol{ \omega} /\gamma$ in Eq.\ \textcolor{blue} {(\ref{LLeq})} is not taken into account, the general expressions for the components and magnitude of the particle angular velocity, \textcolor{blue} {(\ref{nu_3})} and \textcolor{blue} {(\ref{|nu|})}, should be modified by the replacement $\alpha + \beta \to \beta$ (i.e., $q \to \alpha$, $c \to \alpha/\! \sqrt{1 + \alpha^{2}}$, and so on). Since this replacement corresponds to the limiting case $\beta \to \infty$ [see \textcolor{blue} {(\ref{nu_z2})} and \textcolor{blue} {(\ref{|nu|2})}], one can check that the Barnett field does not practically influence the particle rotation at $\beta \gg \alpha$. At the same time, for $\beta \lesssim \alpha$ the difference between the exact results and those obtained by the above replacement is significant [cf.\ \textcolor{blue} {(\ref{beta_to_0})} with \textcolor{blue} {(\ref{nu_z2})} and \textcolor{blue} {(\ref{|nu|2})}]. Therefore, we may conclude that the rotational properties of isotropic ferromagnetic nanoparticles suspended in a viscous liquid and subjected to a rotating magnetic field are determined not only by the Barnett field, but also by the frictional torque [see Eq.\ \textcolor{blue} {(\ref{dyn_omega})}].

\subsection{Numerical results}
\label{NumRes}

For numerical analysis, we use two main assumptions of the model to choose its parameters. The first one was that the time derivative of the particle angular momentum is assumed to be much less than the frictional torque. This assumption, which holds when $\kappa \upsilon \ll \beta$, i.e., $\Omega \ll 60\eta/\rho_{m}d^2$, allowed us to neglect the left-hand side of Eq.\ \textcolor{blue} {(\ref{Dyn:LL1})}. Since in the single-domain state $d< d_{\mathrm{cr}}$ ($d_{\mathrm{cr}}$ is the critical diameter below which this state is realized), the last condition is not too restrictive.

The second assumption was that the nanoparticle material is assumed to be magnetically isotropic. At first sight, according to the definition \textcolor{blue} {(\ref{h_eff})} of the effective magnetic field, this assumption seems to be valid for $h \gg h_{a}$. However, analysis of Eqs.\ \textcolor{blue} {(\ref{Dyn:LL})} shows that it certainly holds if $h \gg h_{a}/ \alpha$ (since, as a rule, $\alpha < 1$, this condition is more strict than $h \gg h_{a}$), i.e., if $H \gg H_{a}/ \alpha$. Because the rotating magnetic field of large amplitude $H$ is difficult to generate, magnetically soft nanoparticles the anisotropy field $H_{a}$ of which is relatively small are most suitable for experimental verification of our predictions.

\begin{figure}[ht]
    \centering
    \includegraphics[totalheight=5.5cm]{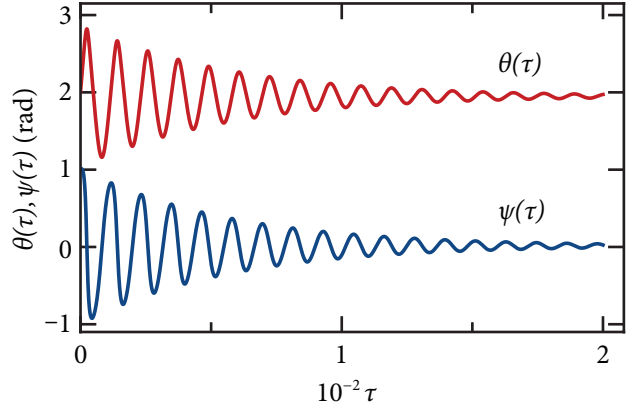}
    \caption{Plots of the functions $\theta =
    \theta(\tau)$ and $\psi = \psi(\tau)$
    obtained via numerical solution of Eqs.\
    \textcolor{blue} {(\ref{DiffEqs})}. The
    parameters and initial conditions are chosen
    to be $q= 0.04$, $h= 0.5$, $\upsilon = 0.2$,
    $\rho=+1$, $\theta(0) = 2\, \mathrm{rad}$
    and $\psi(0) = 1\, \mathrm{rad}$. In the long-time
    limit, the functions $\theta(\tau)$ and
    $\psi(\tau)$ tend to constant values $1.95$
    and $1.48\times 10^{-2}\, \mathrm{rad}$, respectively.
    From \textcolor{blue}
    {(\ref{Sol_rho_1})} it follows that these limiting
    values are in complete agreement  with the analytical
    ones $\theta_{+1}^{(1)}$ and $\psi_{+1}^{(1)}$.}
    \label{fig1}
\end{figure}

As an illustrative example, we consider permalloy nanoparticles ($\mathrm{Ni_{80}Fe_{20}}$) characterized by the parameters \cite{Guim} $M = 8\times 10^{2}\, \mathrm{emu\, cm^{-3}}$, $H_{a} = 4\, \mathrm{Oe}$, $\rho_{m} = 8.7\, \mathrm{g\, cm^{-3}}$, and $\alpha = 0.04$. Note that for these particles $d_{\mathrm{cr}} = 36.8\, \mathrm{nm}$ and $d_{1} = 10.7\, \mathrm{nm}$ at $h=0.1$, i.e., the magnetization fluctuations are negligible if $d_{1} < d < d_{\mathrm{cr}}$. For particles suspended in water at room temperature $T= 298\, \mathrm{K}$ we have $\eta = 8.9 \times 10^{-3}\, \mathrm{P}$, $\beta = 1.18\times 10^{3}$ (we take $\gamma = 1.76 \times 10^{7}\, \mathrm{G^{-1} s^{-1}}$), and so $q=\alpha$ with excellent accuracy. Using these parameters, we numerically solved a set of equations \textcolor{blue} {(\ref{DiffEqs})} for different values of the parameters $h$, $\upsilon$ and $\rho$ controlling the rotating field characteristics. It was established that, according to our predictions, the angles $\theta$ and $\psi$ evolve in an oscillatory manner to the steady-state values \textcolor{blue} {(\ref{Sol_rho_1})} that correspond to the stable magnetization state $\mathbf{m}^{(1)}$. The time dependence of these angles is illustrated in Fig.\ \ref{fig1} for the particular case of the rotating field.

\begin{figure}[ht]
    \centering
    \includegraphics[totalheight=5.5cm]{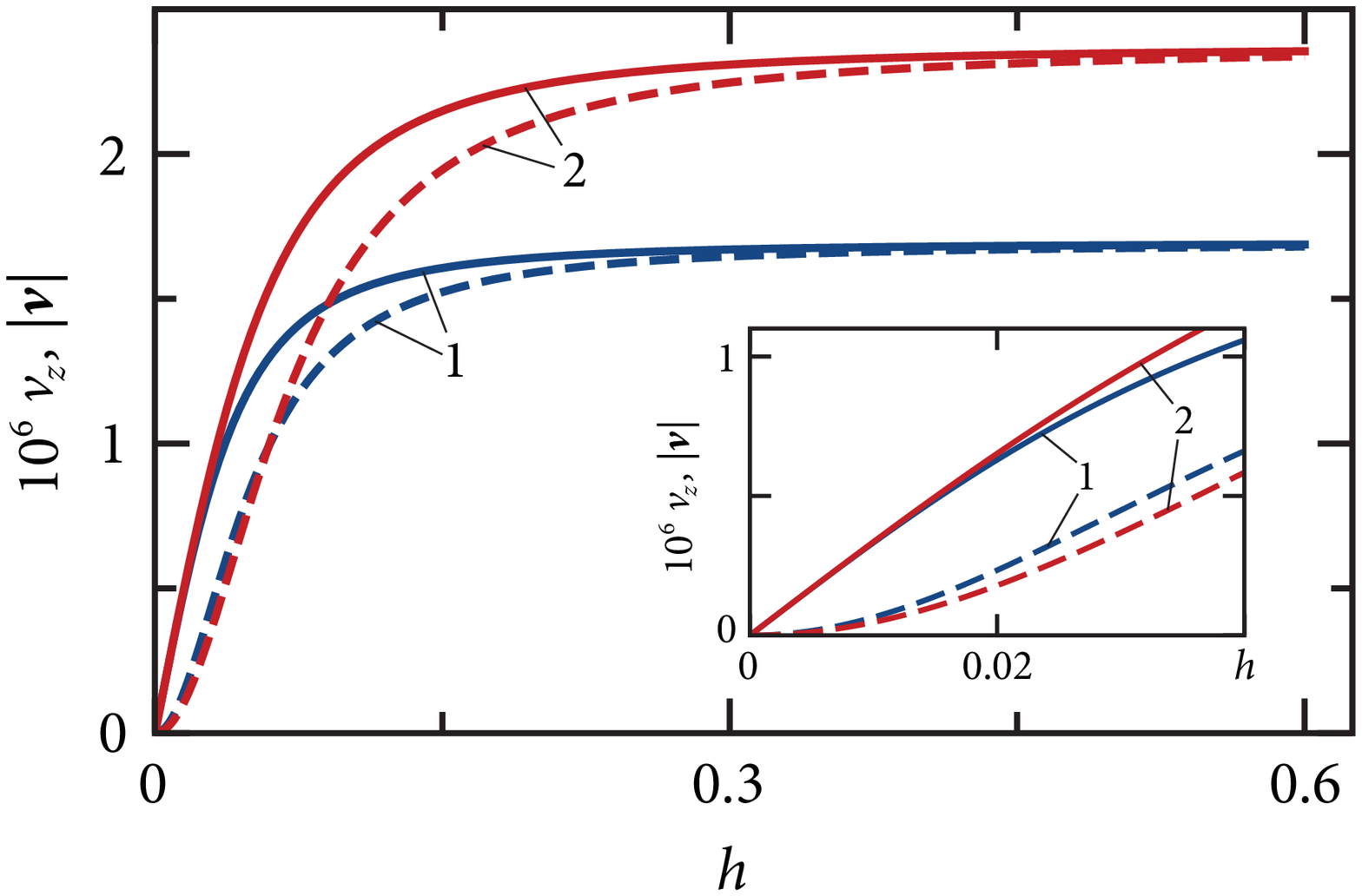}
    \caption{The $z$-components (dashed curves)
    and magnitudes $|\boldsymbol{\nu}|$ (solid
    curves) of the precessional angular velocity
    $\boldsymbol{\nu}$ of permalloy nanoparticles
    as functions of the magnetic field amplitude $h$
    for different values of the magnetic field
    frequency $\upsilon$. Curves 1 correspond to
    $\upsilon = 0.05$ and curves 2 correspond to
    $\upsilon = 0.07$.}
    \label{fig2}
\end{figure}

Dependencies of the $z$-component and magnitude of the precessional angular velocity on the magnetic field amplitude calculated for permalloy nanoparticles from \textcolor{blue} {(\ref{nu_3})} and \textcolor{blue} {(\ref{|nu|})}, respectively, are shown in Fig.\ \ref{fig2}. According to \textcolor{blue} {(\ref{chi_to_inf})}, at small $h$ (when $\chi = \upsilon / h \gg 1$) $\nu_z$ is a quadratic function of $h$, $\nu_z = \alpha h^{2}/ \beta \upsilon$ (we use the relation $q = \alpha$ and assume that $\rho = +1$), and $|\boldsymbol{\nu}|$ grows linearly with $h$, $|\boldsymbol{\nu}| = \alpha h/\beta$ (see inset in Fig.\ \ref{fig2}). If $h$ is large enough (i.e., $\chi \ll 1$), then, using \textcolor{blue} {(\ref{chi_to_0})} and the condition $\alpha^2 \ll 1$, one can make sure that the functions $\nu_z$ and $|\boldsymbol{\nu}|$ approach almost the same value: $\nu_z = |\boldsymbol{\nu}| = \alpha \upsilon/ \beta$. Specifically, $\nu_z = |\boldsymbol{\nu}| = 1.69 \times 10^{-6}$ for $\upsilon = 0.05$ and $\nu_z = |\boldsymbol{\nu}| = 2.37 \times 10^{-6}$ for $\upsilon = 0.07$, or, in dimensional form, $\omega_z = |\boldsymbol{\omega}| = 2.39 \times 10^{4}\, \mathrm{rad\, s^{-1}}$ and $\omega_z = |\boldsymbol{\omega}| = 3.34 \times 10^{4}\, \mathrm{rad\, s^{-1}}$, respectively. Note that, since $(|\boldsymbol{ \nu}| - \nu_z)/ (\alpha \upsilon /\beta) \approx \alpha^{2}/2 \ll 1$, the nanoparticle rotation about the axes $x$ and $y$ is negligibly slow.

\begin{figure}[ht]
    \centering
    \includegraphics[totalheight=5.5cm]{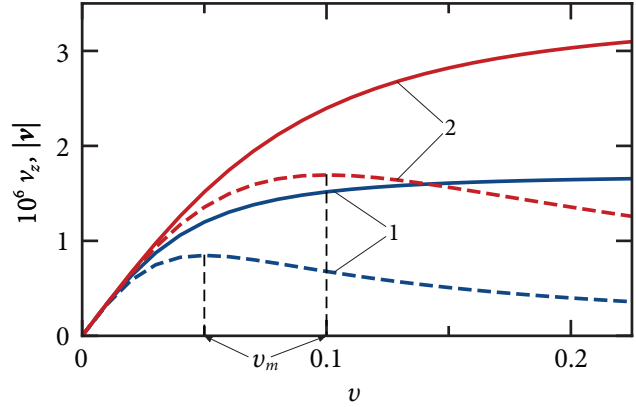}
    \caption{The $z$-components (dashed curves)
    and magnitudes $|\boldsymbol{\nu}|$ (solid
    curves) of the precessional angular velocity
    $\boldsymbol{\nu}$ of permalloy nanoparticles
    as functions of the magnetic field frequency
    $\upsilon$ for different values of the magnetic
    field amplitude $h$. Curves 1 correspond to
    $h = 0.05$ and curves 2 correspond to $h = 0.1$.}
    \label{fig3}
\end{figure}

The behavior of $\nu_z$ and $|\boldsymbol{\nu}|$ as functions of the magnetic field frequency $\upsilon$ is illustrated in Fig.\ \ref{fig3}. If $\upsilon$ is rather small (i.e., $\chi \ll 1$), then, according to \textcolor{blue} {(\ref{chi_to_0})}, $\nu_z$ and $|\boldsymbol{\nu}|$ grow approximately linearly with $\upsilon$: $\nu_z = |\boldsymbol{\nu}| = \alpha \upsilon/ \beta$. In contrast, if $\upsilon$ is rather large (i.e., $\chi \gg 1$), then, according to \textcolor{blue} {(\ref{chi_to_inf})}, $\nu_z$ decreases with $\upsilon$ as $\nu_z = \alpha h^2/ \beta\upsilon$, and $|\boldsymbol{\nu}|$ increases up to $|\boldsymbol{\nu}| = \alpha h/ \beta$. Thus, while $|\boldsymbol{\nu}|$ is a monotonically increasing function of $\upsilon$, $\nu_z$ as a function of $\upsilon$ has a global maximum (recall that $\rho = +1$) at $\upsilon = \upsilon_m$, where $\upsilon_m$ can be estimated as $\upsilon_m \sim h$. In other words, the rotation of isotropic nanoparticles about the $z$ axis, which is induced by the magnetic field of a fixed amplitude $h$ rotating in the $xy$ plane, occurs with the maximal angular velocity, if the magnetic field frequency $\upsilon$ is of the order of $h$ (in dimensional form, if $\Omega \sim \gamma H$).

\begin{figure}[ht]
    \centering
    \includegraphics[totalheight=5.5cm]{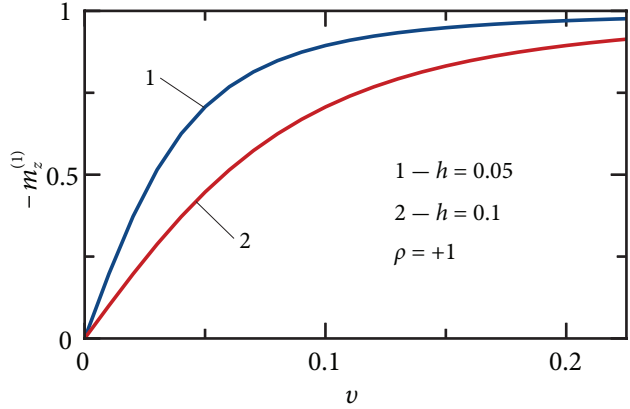}
    \caption{Dependence of the magnetization
    component $-m_{z}^{(1)}$ on the magnetic field
    frequency $\upsilon$ for different values of
    the magnetic field amplitude $h$. }
    \label{fig4}
\end{figure}

It is important to note that, because the magnetization and particle precessions are completely correlated, $\boldsymbol{\nu} \cdot \mathbf{m}^{(1)} = 0$, experimental confirmation of the existence of dissipation-induced rotation of isotropic ferromagnetic nanoparticles could be obtained by analyzing some unique magnetic properties of such systems. In particular, according to \textcolor{blue} {(\ref{m_(l)})}, the $z$ component of the steady-state magnetization $\mathbf{m}^{(1)}$ is given by $m^{(1)}_{z} = - \rho \sqrt{1- R^2/c^2 k^2}$. Using \textcolor{blue} {(\ref{defR})}, it can be shown that $m^{(1)}_{z}$ at $\upsilon \ll h$ is a linear function of $\upsilon$, $m^{(1)}_{z} = - \rho \upsilon /h(1+q^2)$, and $m^{(1)}_{z}$ approaches $-\rho$ at $\upsilon \gg h$ (see Fig.\ \ref{fig4}). The experimental observation of these features would confirm the proposed theory of nanoparticle rotation.

\section{Conclusions}
\label{Concl}

We have predicted and analyzed the precessional rotation of magnetically isotropic ferromagnetic nanoparticles in a viscous liquid generated by a rotating magnetic field. A remarkable feature of this phenomenon is that it occurs when coupling between magnetic and lattice subsystems arising from magnetocrystalline anisotropy is absent. We have shown explicitly that the reason for this rotation is the dissipation-induced coupling between these subsystems.

Our approach to the problem is based on a set of the Landau-Lifshitz equation describing the magnetization dynamics of a magnetically isotropic nanoparticle and the mechanical equation describing the particle rotation in a liquid. Assuming that inertial effects are negligible, we solved these equations analytically and showed that in the steady state both the magnetization and the nanoparticle are precessed. These precessions are fully synchronized and occur about the axis perpendicular to the plane of the magnetic field rotation. We have determined their characteristics and established, in particular, that the precessions occur with the magnetic field frequency. It should be emphasized that, in contrast to an ordinary spinning top, the frequency of particle rotation is much less than the frequency of its precession.

We have also discussed the possibility of experimental detection of the dissipation-induced rotation of isotropic ferromagnetic nanoparticles by a rotating magnetic field. Since direct experimental observation of nanoparticle rotation seems to be problematic, we expect that this phenomenon can be verified by comparing the predicted and experimental magnetic properties of these systems. Such a possibility follows from the strong dissipation-induced coupling between magnetic and lattice subsystems of nanoparticles in the steady state.

\section*{Acknowledgments}

T.V.L.\ and P.H.\ acknowledge the support of the Germany-Ukraine bilateral cooperation project under Deutsche Forschungsgemeinschaft (DFG) Grant No.\ HA 1517/42-1 (P.H.) and Ukrainian State Fund for Fundamental Research (SFFR) Grant No.\ F 81/41894 (T.V.L). In addition, T.V.L.\ and S.I.D.\ are grateful to the Ministry of Education and Science of Ukraine for financial support under Grant No.\ 0119U100772.

\bibliography{DissipationIndRotation}
\end{document}